\documentstyle[preprint,aps]{revtex}
\begin{document}
\draft
\preprint{CLNS 95/1378}
\title{Self Consistent $1/N_c$ Expansion\\
 In The Presence Of Electroweak Interactions}
\author{Chi-Keung Chow and Tung-Mow Yan}
\address{Newman Laboratory of Nuclear Studies, Cornell University, Ithaca,
NY 14853.}
\date{\today}
\maketitle
\begin{abstract}
In the conventional approach to the $1/N_c$ expansion, electroweak
interactions are switched off and large $N_c$ QCD is treated in isolation.
We study the self-consistency of taking the large $N_c$ limit in the presence
of electroweak interaction.
If the electroweak coupling constants are held constant, the large $N_c$
counting rules are violated by processes involving internal photon or weak
boson lines.
Anomaly cancellations, however, fix the ratio of electric charges of
different fermions.
This allows a self-consistent way to scale down the electronic charge $e$
in the large $N_c$ limit and hence restoring the validity of the large $N_c$
counting rules.
\end{abstract}
\narrowtext
The $1/N_c$ expansion is now generally recognized as an invaluable tool in
our handling of the non-perturbative nature of hadron dynamics.
The pioneer work of 't~Hooft \cite{1} has proven that the large $N_c$
limit is the weak coupling limit of meson dynamics.
Quantitatively, a graph with $k$ external meson legs can be {\it at most\/}
of order $N_c^{1-k/2}$.
For example, meson masses, described by graphs with two external meson
legs, are of order $N_c^0$, while the ``meson $\to$ meson + meson'' decay
amplitudes are suppressed by $N_c^{-1/2}$.
As a result, mesons are stable and non-interacting in the large $N_c$
limit.
The generalization to include baryons was made by Witten \cite{2}, who had
shown that a graph with two external baryon legs and $k$ external meson legs
are {\it at most\/} of order $N_c^{1-k/2}$.
Hence the baryon masses ($k=0$) and Yukawa couplings ($k=1$) grow like
$N_c$ and $N_c^{1/2}$ respectively.

In the real world, however, hadrons experience not only QCD but also
electroweak interactions.
In the conventional approach to the $1/N_c$ expansion, one simply ignores
the electroweak interactions and treats large $N_c$ QCD in isolation.
The results obtained for large $N_c$ are then extrapolated back to $N_c=3$
and applied to electroweak processes.
This practice is permissible since electroweak coupling constants are
independent parameters.
It is interesting to ask what happens to the large $N_c$ counting rules
described above if the electroweak interactions are not switched off\footnote
{We believe we are not the first ones to raise this question.
In Chapter~7 of Ref.~\cite{2.5} Marshak made a cautionary remark about taking
the large $N_c$ limit ``when the leptonic and quark sectors are both involved
in the process'' (pg.~450).
We are, however, not aware of any systematic discussion on this topic in
the literature.}.
If the electroweak theory is not modified, it is easy to see that these
counting rules will be violated by graphs involving electroweak currents.

One possible violation is the $\rho$ meson two point function induced by
$\rho$--$\gamma$ mixing.
As mentioned above, the $\rho$ meson mass should be of order $N_c^0$.
The $\rho$--$\gamma$ mixing parameter, however, is just governed by the
$\rho$ meson decay constant $f_\rho$, which grows like
$N_c^{1/2}$.
It follows that the contribution to the $\rho$ mass by the
$\rho$--$\gamma$--$\rho$ mixing diagram grows like
\begin{equation}
m_\rho \sim f_\rho^2 \sim N_c^1,
\end{equation}
violating the counting rule above.
Another example is the $\pi^0 \to 2\gamma$ decay amplitude by the
Adler--Bell--Jackiw anomaly \cite{3,4},
\begin{equation}
A_{\pi^0} \sim N_c / f_\pi \sim N_c^{1/2},
\end{equation}
which diverges in the large $N_c$ limit, in contradiction with the claim
of meson stability made above.
Moreover, such $\pi^0\gamma\gamma$ vertices can induced large $\pi^0\pi^0$
elastic scattering (through photon loops) amplitude of order $N_c^2$,
violating the counting rule requirement that meson-meson elastic scattering
amplitude should decrease like $N_c^{-1}$.

Such violations are also present in the baryon sector.
Consider baryons with $N_c$ quarks with the same flavor and hence the
same electric charge.
(For up and down quarks they are the large $N_c$ generalizations of the
$\Delta^{++}$ and $\Delta^-$ baryons respectively.)
The electrostatic energies carried by such baryons grow like $N_c^2$, in
violation of the counting rule that baryon masses should grow like $N_c^1$
only.
These examples of violations of large $N_c$ counting rules reflect the
unsmoothness of the large $N_c$ limit in the presence of electroweak
interactions.
Electromagnetic interactions introduce a correction of relative order
$e^2N_c$ to some strong processes.
These effects vanish if we set $e^2=0$, but otherwise they diverge in the
large $N_c$ limit, independent of the particular values taken by $e$.

In this paper, we will try to show that the success of the $1/N_c$
expansion is no accident.
There exists a well-behaved large $N_c$ limit even in the presence of
electroweak interactions.
We will show that the ratio of electric charges carried by quarks and
leptons are fixed by anomaly cancellation \cite{5,6} of the underlying
SU($N_c$)$_c\,\times\,$SU(2)$_L\,\times\,$U(1)$_Y$ gauge theory.
To achieve a smooth large $N_c$ limit one can consistently scale down the
electric charges carried by all the particles by a common power of $N_c$.
This will introduce extra powers of $1/N_c$ to graphs involving photon
currents and keep them in agreement with the large $N_c$ counting rules.
In addition, the modified electroweak interactions will remain a small
perturbation to QCD, as they are in the real world.

For the SU($N_c$)$_c\,\times\,$SU(2)$_L\,\times\,$U(1)$_Y$ gauge theory to
be renormalizable, it is necessary to have all chiral anomalies cancelled.
For example, the triangular anomalies \cite{7}, describing the interaction of
three gauge bosons interaction through fermion loops, must be cancelled
exactly.
In one generation standard model, the fermions fall into the representations
listed below:

\medskip
\centerline{\vbox{
\halign{\hfil#&\qquad\qquad\hfil#\hfil&\qquad\hfil#\hfil&\qquad\hfil#\hfil\cr
Fields&SU$(N_c)_c$&SU(2)$_L$&U(1)$_Y$\cr
$u_L \choose d_L$&3&2&$Y_Q$\cr
$u_R\phantom)$&3&1&$Y_u$\cr
$d_R\phantom)$&3&1&$Y_d$\cr
$\nu_L \choose e_L$&1&2&$Y_L$\cr
$e_R\phantom)$&1&1&$Y_e$\cr}}}
\medskip

\widetext
Only the following triangular anomalies do not cancel trivially and provide
constraints on the hypercharges of different fermions.
\begin{mathletters}
\begin{eqnarray}
\hbox{U(1)}_Y^3:&
\qquad 2N_c Y_Q^3 - N_c Y_u^3 - N_c Y_d^3 + 2 Y_L^3 - Y_e^3 = 0,
\label{a}\\
\hbox{U(1)}_Y\hbox{SU(2)}_L^2:&\qquad N_c Y_Q + Y_L = 0,
\label{b}\\
\hbox{U(1)}_Y\hbox{SU}(N_c)_c^2:&\qquad 2 Y_Q -Y_u -Y_d = 0,
\label{c}\\
\noalign{\medskip\hbox{
and the mixed gauge-gravitational anomaly \cite{8,9,10} provide a fourth
constraint\footnote{Yet another chiral anomaly, the global chiral SU(2)
anomaly \cite{11}, constrain the number of left-handed fermion doublets to be
even, hence requiring $N_c$ to be odd and leaving the baryons as fermions.}: }
\medskip}
\hbox{U(1)}_Y\hbox{(graviton)}^2:&
\qquad 2N_c Y_Q - N_c Y_u - N_c Y_d + 2 Y_L - Y_e = 0.
\label{d}
\end{eqnarray}
\end{mathletters}
One can eliminate $Y_L$ and $Y_e$ from Eq.~(\ref{a}) by Eq.~(\ref{b}) and
Eq.~(\ref{d}).
With $Y = {1\over2}(Y_u - Y_d)$, Eq.~(\ref{c}) gives $Y_u = Y_Q + Y$ and
$Y_d = Y_Q - Y$, and Eq.~(\ref{a}) becomes
\begin{equation}
2N_c Y_Q^3 - N_c (Y_Q + Y)^3 - N_c (Y_Q - Y)^3 + 2 (-N_c Y_Q)^3 - (-2N_c Y_Q)^3
= 0,
\end{equation}
which can be further reduced to
\begin{equation}
Y_Q(N_c^2 Y_Q^2 - Y^2) = 0.
\end{equation}
There are clearly two solutions to this equation.
The ``bizarre'' solution \cite{12} with $Y_Q = 0$ which gives
\begin{equation}
Y_Q = Y_L = Y_e = 0,\qquad Y_u = -Y_d,
\end{equation}
is phenomenologically uninteresting for reasons detailed in Ref.~\cite{13}.
That leaves us with the ``standard'' solution with $Y = N_c Y_Q$ (choosing
$Y = -N_c Y_Q$ just reverses the labels ``up'' and ``down'' quarks),
\begin{equation}
(Y_Q, Y_u, Y_d, Y_L, Y_e) = (1, N_c+1, -N_c+1, -N_c, -2N_c) Y_Q.
\end{equation}
All the hyperchrages are fixed up to an overall proportionality constant.
\narrowtext

The electric charge is defined as,
\begin{equation}
Q = e(I_3 + Y/Y_0).
\end{equation}
Since electromagnetic interactions conserves parity, the left-handed quarks
and leptons must carry the same electric charges as their right-handed
counterparts.
This fixes $Y_0 = 2N_c Y_Q$ and
\begin{equation}
(Q_u, Q_d, Q_e, Q_\nu) = ({N_c+1\over 2N_c}, {-N_c+1\over 2N_c}, -1, 0)e.
\label{q}
\end{equation}
By putting $N_c=3$, the normal charge assignments are recovered.
Hence we have shown that charge quantization follows from anomaly
cancellations for arbitrary odd $N_c$.
This observation is crucial for our later discussion as it provides a unique
way to scale down all the charges of the quarks by scaling down the electronic
charge $e$ with anomaly cancellation all the way.

The world described by Eq.~(\ref{q}) shares many features of the real world.
The neutrino is still electrically neutral, and the $Q_u-Q_d=-Q_e$ equality
is preserved so that $\beta$-decays can still happen.
In the large $N_c$ limit, the up and down quarks carry charges $+e/2$ and
$-e/2$ respectively ($2e/3$ and $-e/3$ in the real world), but the $q\bar q$
mesons still have charges $e$, 0, or $-e$ as in the real world.
The proton has $(N_c+1)/2$ up quarks and $(N_c-1)/2$ down quarks and hence
its charge is
\begin{equation}
Q_p = e\left({N_c+1\over2}{N_c+1\over2N_c} - {N_c-1\over2}{N_c-1\over2N_c}
\right) = e,
\end{equation}
and the hydrogen atom stays neutral.
The neutron, on the other hand, carries no electric charge as usual.
\begin{equation}
Q_n = e\left({N_c-1\over2}{N_c+1\over2N_c} - {N_c+1\over2}{N_c-1\over2N_c}
\right) = 0.
\end{equation}

Coming back to large $N_c$ counting rules, the $\pi^0 \to 2\gamma$ decay
is given by,
\begin{equation}
A_\pi^0 \sim {N_c(Q_u^2-Q_d^2)\over f_\pi}.
\end{equation}
As mentioned above, $N_c/f_\pi \sim N_c^{1/2}$ but now we have an additional
suppression factor from the electric charges, $Q_u^2-Q_d^2 = e^2/N_c$.
Hence
\begin{equation}
A_\pi^0 \sim {e^2 \over f_\pi} \sim N_c^{1/2},
\end{equation}
and the counting rules are satisfied.

The $\rho$--$\gamma$ mixing problem, however, still persists.
The $\rho$--$\gamma$ mixing amplitude $A_{\rho\gamma}$ is given by,
\begin{equation}
A_{\rho\gamma} = f_\rho (Q_u-Q_d) = ef_\rho,
\end{equation}
which diverges as before.
Also, the $\Delta$ baryon self-energy still diverges as $N_c^2$, violating the
counting rules.

As suggested before, one of the possible remedies to the situation is to
scale down the electronic charge $e$ in the large $N_c$ limit,
providing extra suppression factors.
Since we are scaling the strong coupling constant $g_3$ by keeping $g_3^2N_c =
\hbox{constant}$, it is natural to impose the electric charge scaling
condition as
\begin{equation}
e^2 N_c = \hbox{ constant, \qquad as } N_c \to \infty.
\label{cond}
\end{equation}
With $e^2 \sim N_c^{-1}$, $A_{\rho\gamma}$ is suppressed in the large $N_c$
limit,
\begin{equation}
A_{\rho\gamma} = ef_\rho \sim N_c^0,
\end{equation}
and the $\Delta$ baryon electrostatic self energy is
\begin{equation}
M_{elec} \sim e^2 N_c^2 \sim N_c,
\end{equation}
exactly as specified by the counting rules.
In general, it is easy to prove that Eq.~(\ref{cond}) is sufficient to keep
all the large $N_c$ counting rules intact even in the presence of photons.
We first note that the $q\bar q\gamma$ vertex is of the same order as the
$q\bar qg$ vertex (both of order $N_c^{-1/2}$), and replacing an internal
gluon line from a planar diagram with a photon line does not produce
additional powers of $N_c$.
An analysis similar to the one given by Witten \cite{2} can be readily
carried out.
Moreover, the couplings of quarks to leptons or $W^\pm$ via exchange of
photons present no difficulties as a consequence of Eq.~(\ref{cond}).
Thus the graphs with photon lines are either of the same order in $N_c$ as
the leading planar diagrams, or are simply dominated by the latter.
Hence condition~(\ref{cond}) is sufficient to guarantee the validity of the
large $N_c$ counting rules.

Our conclusions can be easily generalized to the case of weak currents.
The graphs with $W^\pm$ and $Z^0$ lines can violate the large $N_c$ counting
rules unless the conditions like
\begin{equation}
g_2^2 N_c = \hbox{ constant, \qquad as }N_c \to\infty,
\label{cond2}
\end{equation}
are imposed, where $g_2$ is the SU(2)$_L$ coupling constant.
It is a general feature that {\it all\/} coupling constants must be scaled
down correspondingly even though we are taking the large $N_c$ limit of
only {\it one\/} of the gauge groups.

One should also note that the large $N_c$ scaling conditions Eq.~(\ref{cond})
and (\ref{cond2}) are not the only ones which lead to a smooth large $N_c$
limit.
It is easy to see that any scaling conditions like
\begin{equation}
e^2 \sim N_c^{-m}, \hbox{\qquad as }N_c \to \infty,
\end{equation}
with $m\geq 1$ is going to give a smooth $1/N_c$ limit.
Conditions~(\ref{cond}) and (\ref{cond2}) are just the critical cases with
$m=1$.
A large suppression power $m$, on the other hand, will lead to severe
suppression of electroweak effects in the large $N_c$ limit.
It is noted that the conventional approach of switching off the electroweak
interaction before taking the large $N_c$ limit is equivalent to taking
$m \to \infty$ in this formalism.

Lastly, it is natural to ask if there exists any well-defined limit of the
weak mixing angle $\theta_W$ in the large $N_c$ limit.
It seems to us that, since the coupling constants of U(1)$_Y$ and SU(2)$_L$
are independent quantities, the value of $\theta_W$ is not constrained
unless we start with some grand unified gauge group.
It turns out that no simple analogs of SU(5) or SO(10) grand unifications
exist for SU($N_c)_c\,\times\,$SU(2)$_L\,\times\,$U(1)$_Y$.
Hence $\theta_W$ is unconstrained in the present stage of our understanding
of the large $N_c$ limit.

\acknowledgements
C.K.C. would like to thank Dan Pirjol for asking about the $\pi^0\to 2\gamma$
process in the large $N_c$ limit, which initiates this study.
This work is supported in part by the National Science Foundation.

\end{document}